%% file: acl_latex.tex
\pdfoutput=1

\documentclass[11pt,table]{article}

\usepackage[final]{acl}

\usepackage{times}
\usepackage{latexsym}
\usepackage{fontawesome5} 
\usepackage{tikz}
\usepackage{amsmath}
\usepackage{booktabs}
\usepackage{wrapfig}

\newenvironment{AIbox}[1]{%
    \begin{tcolorbox}[
        colframe=black!20!white,
        colback=black!5!white,
        boxrule=0.5pt,
        arc=2mm,
        title=#1,
        fonttitle=\bfseries
    ]
}{%
    \end{tcolorbox}
}

\usepackage[utf8]{inputenc} 
\usepackage[T1]{fontenc}    
\usepackage{hyperref}       
\usepackage{url}            
\usepackage{booktabs}       
\usepackage{amsfonts}       
\usepackage{array}          
\usepackage{nicefrac}       
\usepackage{microtype}      
\usepackage{booktabs} 
\usepackage{multirow} 
\usepackage{xspace}
\usepackage{multirow}
\usepackage{booktabs} 
\usepackage{caption}  
\usepackage{graphicx}
\usepackage{bbding}
\usepackage{tcolorbox}
\usepackage{textcomp}
\newcommand{\fname}{Humanity's Last Code Exam\xspace}
\newcommand{\name}{HLCE\xspace}
\usepackage{tikz}
\usepackage[T1]{fontenc}
\newcommand{\progressbar}[4]{%
  \begin{tikzpicture}[baseline]
    \def\ratio{#1}
    \def\w{#2}\def\h{#3}
    \draw[rounded corners=1pt] (0,0) rectangle (\w,\h);
    \fill[#4,rounded corners=1pt] (0,0) rectangle ({\ratio*\w},\h);
  \end{tikzpicture}%
}

\usepackage{subcaption}

\usepackage[utf8]{inputenc}
\usepackage{microtype}
\usepackage{inconsolata}
\usepackage{graphicx}
\usepackage{enumitem}
\usepackage{hyperref} 

\title{
  \begin{minipage}{0.80\textwidth}
    \raisebox{-0.4\height}[0pt][0pt]{
      \includegraphics[height=3.0\baselineskip]{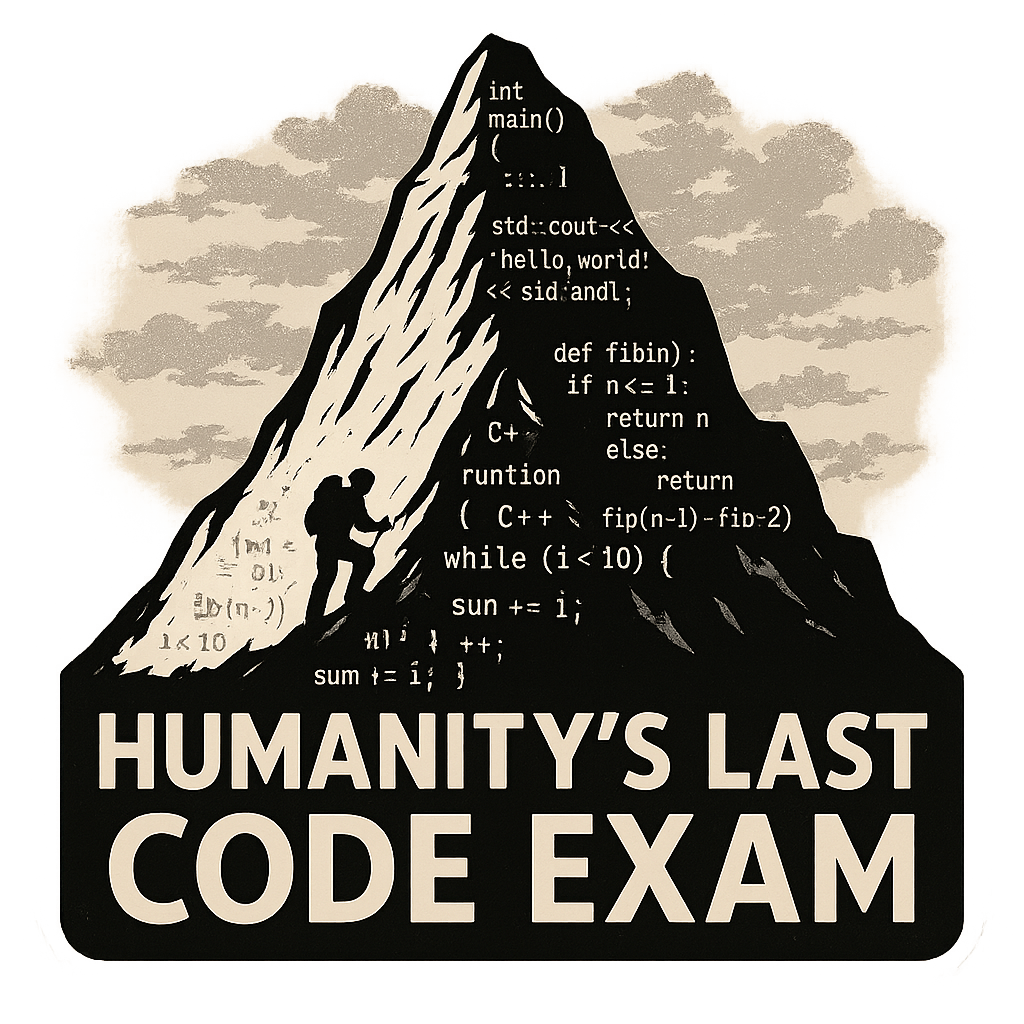}}
      \hspace{-5pt}
    \begin{minipage}{1.0\textwidth}
    \centering
       Humanity's Last Code Exam: \\  Can Advanced LLMs Conquer Human's  Hardest Code Competition?
    \end{minipage}
  \end{minipage}
}

\author{%
  Xiangyang Li\thanks{These authors contributed equally to this work.} \quad Xiaopeng Li\footnotemark[1] \quad Kuicai Dong \quad Quanhu Zhang \quad \\
  \textbf{Rongju Ruan} \quad \textbf{Xinyi Dai}\thanks{Corresponding authors.} \quad \textbf{Xiaoshuang Liu} \quad \textbf{Shengchun Xu} \quad \\
  \textbf{Yasheng Wang} \quad \textbf{Ruiming Tang}\footnotemark[2] \\
  \textsc{Huawei Noah's Ark Lab} \\
  \texttt{\{lixiangyang34, dong.kuicai, zhangquanhu1, ruanrongju, daixinyi5, } \\ 
  \texttt{liuxiaoshuang4, xushengchun, wangyasheng, tangruiming\}@huawei.com} \\
  \texttt{xiaopli2-c@my.cityu.edu.hk}
}
\begin{document}

\maketitle
\begin{abstract}
Code generation is a core capability of large language models (LLMs), yet mainstream benchmarks (e.g., APPs and LiveCodeBench) contain questions with medium-level difficulty and pose no challenge to advanced LLMs. To better reflected the advanced reasoning and code generation ability,  We introduce \textbf{\underline{H}}umanity’s \textbf{\underline{L}}ast \textbf{\underline{C}}ode \textbf{\underline{E}}xam (\textbf{HLCE}), comprising 235 most challenging problems
from the International Collegiate Programming Contest (ICPC World Finals) and the International Olympiad in Informatics (IOI) spanning 2010 – 2024.
As part of HLCE, we design
a harmonized online–offline sandbox that guarantees fully reproducible evaluation. Through our comprehensive evaluation, we observe that even the strongest reasoning LLMs: o4-mini(high) and Gemini-2.5 Pro, achieve pass@1 rates of only 15.9\% and 11.4\%, respectively. Meanwhile, we propose a novel "self-recognition" task to measure LLMs' awareness of their own capabilities. Results indicate that LLMs' self-recognition abilities are not proportionally correlated with their code generation performance. Finally, our empirical validation of test-time scaling laws reveals that current advanced LLMs have substantial room for improvement on complex programming tasks.
 We expect HLCE to become a milestone challenge for code generation and to catalyze advances in high-performance reasoning and human–AI collaborative programming. Our code and dataset are also public available\footnote{\url{https://github.com/Humanity-s-Last-Code-Exam/HLCE}}.

\end{abstract}

\input{tables/model_overview}
\input{tex/1introduction}

\input{tex/2related}
\input{tex/3benchmark}
\input{tex/4experiments}
\input{tex/5dicussion}

\section{Conclusion}
In this work, we introduced \name, a challenging benchmark of 235 competitive programming problems from IOI and ICPC World Finals. Top models achieve only 15.1\% and 11.4\% pass@1 rates. Our benchmark includes standard and interactive programming challenges alongside a novel self-assessment task. By incorporating human competition data, we established metrics comparing LLMs with top-tier programmers, revealing substantial room for improvement. Test-time scaling law validation confirms current models have not reached performance ceilings, suggesting promising directions for advancing LLMs' reasoning in complex programming tasks. \name aims to drive progress toward code LLMs that  reach the proficiency level of elite human competitors.

\section{Limitations}
 Due to API pricing constraints and inference latency limitations, we could only generate 5 responses for each reasoning model. With more responses, LLMs would likely achieve better results in historical competitions. However, such a comparison might not be entirely fair, as within the specified competition time frame, human coding speed cannot match that of LLMs.

 IOI problems require online submissions, resulting in longer submission processing times—approximately three minutes per problem on average. While this delay is generally acceptable for our evaluation purposes, it does add operational overhead to the   evaluation process.

\section{Ethical Considerations}
We ensure that the distribution of each dataset complies with the corresponding licenses, all of which are listed below:
\begin{enumerate}
    \item[$\bullet$] IOI: Provided under ``CC-BY-SA 4.0'' license.
    \item[$\bullet$] ICPC World Finals: Provided under ``CC-BY-SA 4.0'' license.

\end{enumerate}

For the new artifacts contributed in \name, including but not limited to the questions, test cases, and evaluation scripts, we make them available solely for research purposes. Users are permitted to use, modify, and share these annotations for academic and non-commercial research activities. Any other use, including commercial exploitation, is not permitted without explicit written permission from the authors.

\bibliography{custom}

\appendix

\section{Appendix}
\label{sec:appendix}
\subsection{Dataset Statistic}
The total quantity of filtered \name is presented in Table~\ref{tab:competition_counts}.

\begin{table}[h]
\centering
\begin{tabular}{ccc}
\hline
\textbf{Year} & \textbf{ICPC World Finals } & \textbf{IOI } \\
\hline
2010 & - & 8 \\
2011 & 11 & 6 \\
2012 & 12 & 6 \\
2013 & 11 & 6 \\
2014 & 12 & 6 \\
2015 & 13 & 6 \\
2016 & 13 & 6 \\
2017 & 12 & 5 \\
2018 & 5 & 6 \\
2019 & 11 & 5 \\
2020 & 14 & 6 \\
2021 & 12 & 6 \\
2022 & 10 & 6 \\
2023 & 10 & 6 \\
2024 & - & 6 \\
\hline
\end{tabular}
\caption{Problem Counts in ICPC and IOI by Year}
\label{tab:competition_counts}
\end{table}

\subsection{Problem Example}
In Figures~\ref{fig:icpc_problem} and~\ref{fig:ioi_problem}, we present sample problems from IOI and ICPC World Finals respectively.
\input{figures/icpc_question_response}
\input{figures/IOI_problems}
\subsection{API Cost}
\label{sec:api_cost}
\input{tables/api_cost}

We calculated the actual API costs incurred during our evaluation process, with results presented in Table~\ref{tab:api_cost}. Interestingly, the o4-mini model demonstrated remarkably low cost, emerging as the most economical among all inference models evaluated. This finding is particularly significant as it suggests a promising development trend for code-oriented LLMs: substantial performance improvements may be achievable with minimal financial investment.

The cost-effectiveness of the o4-mini model indicates that efficient architectures and training methodologies can significantly reduce computational expenses without compromising performance quality. This balance between cost and capability represents an important direction for the future development of code generation and inference systems, potentially democratizing access to powerful code LLMs for a wider range of applications and users.

\subsection{Evaluation Details}
\subsubsection{Evaluation Parameters}
For ICPC World Finals problems, following the LiveCodeBench, we set the execution timeout to 30 seconds without imposing memory limitations. For IOI problems, we utilize the web response information returned by the bot.

\subsubsection{Used Prompt}
In this section, we present all the task prompts utilized for the ICPC and IOI competitions, as illustrated in Figures~\ref{fig:pr},~\ref{fig:icpc_generation_prompts}, and~\ref{fig:prompt_self_reco}.
\input{figures/IOI_prompt}
\input{figures/ICPC_world_finals}
\input{figures/self-recog}

\subsubsection{Evation Metric}
\paragraph{Pass@K} 
To evaluate the performance of code generation models, we employ the pass@k metric. This metric assesses a model's ability to generate at least one correct solution within k attempts.

Given a coding problem, we sample k candidate solutions from the model and evaluate each solution against test cases. The pass@k metric is defined as:

\begin{equation}
\text{pass@k} = \mathbb{E}_{x \sim D}\left[1 - \frac{\binom{n-c}{k}}{\binom{n}{k}}\right]
\end{equation}

where:
\begin{itemize}
    \item $x$ represents a coding problem sampled from dataset $D$
    \item $n$ is the total number of solutions generated for each problem
    \item $c$ is the number of correct solutions among the $n$ generations
    \item $\binom{n-c}{k}$ and $\binom{n}{k}$ represent binomial coefficients
\end{itemize}

Intuitively, this formula calculates the probability of finding at least one correct solution when randomly selecting k samples from n generated solutions. When $n = k$, the formula simplifies to $c/n$, which is the fraction of correct solutions. For our experiments, we report pass@1, pass@5 to provide a comprehensive view of model performance across different sampling scenarios.

\paragraph{AUC (Area Under the Curve)} 
The Area Under the Curve (AUC) measures the performance of binary classification models by quantifying the area under the Receiver Operating Characteristic (ROC) curve. Mathematically, AUC is defined as:

\begin{equation}
AUC = \int_{0}^{1} TPR(FPR^{-1}(t)) dt
\end{equation}

where TPR is the True Positive Rate and FPR is the False Positive Rate.

In practice, AUC is computed using the trapezoidal rule:


\begin{align}
AUC \approx \frac{1}{2} \sum_{i=1}^{n-1}  \Big( \ & (FPR_{i+1} - FPR_i) \notag \\
                                            & \cdot (TPR_i + TPR_{i+1}) \Big)
\end{align}

The AUC represents the probability that a randomly selected positive instance will rank higher than a randomly selected negative instance. An AUC of 1.0 indicates perfect classification, while 0.5 suggests performance equivalent to random chance.

In the self-recognition task, we employ the AUC metric to evaluate whether LLMs can accurately identify if their generated answers are correct or incorrect. This metric provides a comprehensive measure of the model's ability to discriminate between its own correct and incorrect responses, effectively quantifying the model's self-recognition capabilities.


\end{document}

%% file: tables/model_overview.tex
\begin{figure*}[ht]
    \centering
    \includegraphics[width=0.98\linewidth]{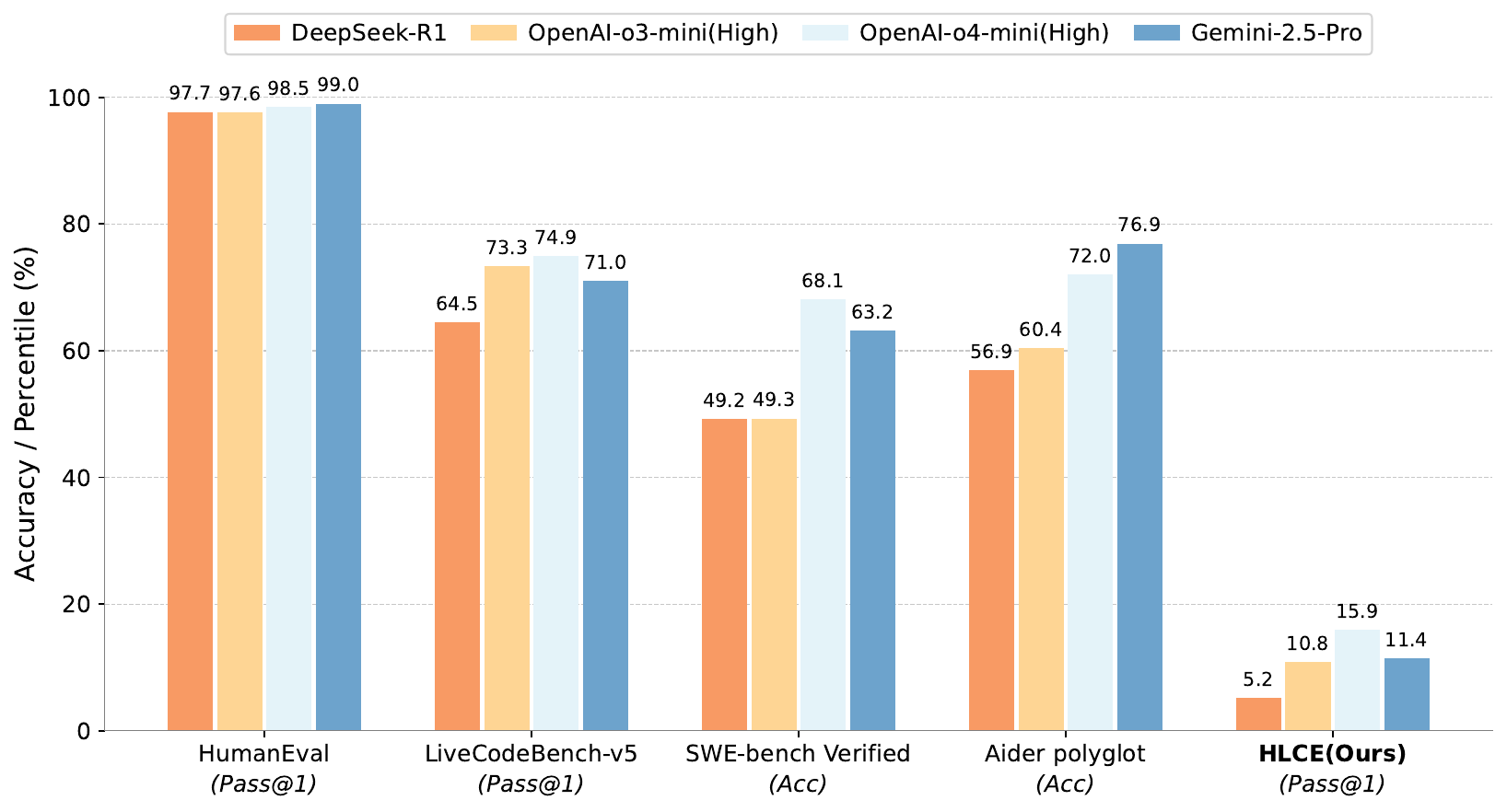}
    \vspace{-1em}
    \caption{Performance of advanced LLMs on widely-used code generation benchmarks and \name.}
    \label{fig:overview}
\end{figure*}

%% file: tex/1introduction.tex
\section{Introduction}

Large Language Models (LLMs)~\cite{deepseek,gpt4,o1,qwen2.5coder} have demonstrated human-level proficiency across a wide range of text understanding~\cite{dong-etal-2023-open}, reasoning~\cite{yang2025llmreason}, and generation tasks. Among these, code understanding and generation~\cite{qwen2.5coder,li-etal-2025-coir, deepseek-coder,deepseek-coder-v2} have emerged as key research areas, since code inherently reflects reasoning and logical thinking skills. To evaluate LLMs’ capabilities in this domain, many code generation benchmarks, such as LiveCodeBench~\cite{livecodebench}, MBPP~\cite{mbpp}, HumanEval~\cite{humaneval}, and SWE-bench~\cite{swe-bench}, have been established. These benchmarks have accelerated the progress of LLM-based code models, but the most advanced LLMs are now achieving near-perfect or saturated performance on many of these tasks.
This raises the question: Do current benchmarks truly reflect the advanced reasoning and code generation abilities of state-of-the-art LLMs, especially when it comes to complex, structured, or algorithmic coding challenges?

Current benchmarks suffer from several critical limitations:  
\textbf{(1) Limited Difficulty:} With the increasing capabilities of LLMs, many benchmarks have become too easy, as shown in Figure~\ref{fig:overview}. For example, benchmarks like HumanEval already report high pass rates, and achieving perfect scores is now a matter of incremental progress.  
\textbf{(2) Absence of Interactive-based Evaluation:} Most contemporary code competition benchmarks (e.g., LiveCodeBench) rely on standard I/O judging, where code submissions are evaluated via input/output pairs. This facilitates outcome-reward reinforcement learning~\cite{deepseek-r1,qwen3, duong2025improving}, allowing models to exploit these feedback loops. However, interactive-based judging\footnote{Interactive-based judging is such as International Olympiad in Informatics (IOI) problems, where participants implement specific function signatures and the evaluation system interacts with these functions directly} remains largely unexplored for LLMs. The abilities of LLMs in this setting are thus insufficiently characterized.  
\textbf{(3) Underexplored Test-time Scaling Laws:} The relationship between model scale and performance at test-time in code generation is insufficiently probed. While recent models like o4-mini and Gemini-2.5-pro have achieved impressive results, the scalability and potential upper bounds of test-time performance are still open questions that current benchmarks do not address.

To address these limitations, we introduce a new benchmark, \fname (\name), comprising carefully curated and rigorously filtered programming problems sourced from the IOI and ICPC World Finals competitions from 2010 to 2024. After extensive manual cleaning and validation, the benchmark contains 235 high-quality competitive programming problems, each accompanied by comprehensive test cases. HLCE extends beyond traditional ACM input/output formats to include interactive challenges that demand dynamic program behavior and real-time interaction.

Using HLCE, we evaluate 12 leading LLMs, including both reasoning and non-reasoning models. Our results indicate that even state-of-the-art models o4-mini(high) and Gemini-2.5 Pro achieve only 15.1\% and 11.4\% pass@1 rates, respectively, on HLCE. This demonstrates that HLCE presents a significantly higher level of difficulty compared to existing code generation benchmarks. Furthermore, we introduce a novel self-recognition task, in which LLMs must determine whether their generated code solutions are correct or incorrect, providing direct insight into the models’ capacity for self-assessment and error recognition.
Additionally, HLCE enables empirical analysis of test-time scaling laws. Our findings reveal that, despite the impressive capabilities of models like o4-mini(high) and Gemini-2.5 Pro, the upper bounds of test-time performance remain largely unexplored, suggesting considerable room for LLM-based reasoning.

Finally, we consider an intriguing question: Can advanced LLMs genuinely compete with top human programmers in IOI and ICPC World Finals? To explore this, we collected performance data for medalists from these competitions spanning 2010 to 2024 and directly compared their results to those of leading LLMs. Our analysis shows that models such as o3-mini, Gemini-2.5-pro, and o4-mini are capable of achieving medal-level performance in ICPC competitions. However, these models still underperform compared to human medalists in the IOI, highlighting the persisting challenges in truly  top-tier human-competitive code generation.
In summary, our contributions are as follows:
\begin{itemize}[leftmargin=*, itemsep=0.0em, topsep=0.0em]
    \item We introduce HLCE, a novel benchmark comprising 235 competitive programming problems from IOI and ICPC World Finals (2010-2024), featuring both standard and interactive programming challenges that significantly exceed the difficulty of existing code generation benchmarks.
    
    \item We conduct comprehensive evaluations on 12 leading LLMs, showing that even most advanced LLMs achieve only 15.1\% and 11.4\% pass@1 rates.
    We also propose a novel self-recognition task to measure models' abilities to recognize the correctness of their own generated solutions.

    \item We empirically validate test-time scaling laws on HLCE, demonstrating that current LLMs have not yet reached their performance upper bounds and highlighting substantial room for further advancement via improved reasoning capabilities.
    
    \item We provide comparative analyses with top human competitors, revealing the gap between advanced LLMs and competition medalists.

\end{itemize}

%% file: tex/2related.tex
\section{Related Work}

\input{tables/dataset_comparison}

\textbf{Large Language Models for Code Generation.} Recent advancements in LLMs have significantly enhanced the capability of automated code generation. Models such as Codex \cite{humaneval}, StarCoder \cite{starcoder}, and CodeLlama \cite{codellama} have demonstrated remarkable proficiency in understanding and generating programming code across various languages. The emergence of instruction-tuned models like ChatGPT \cite{gpt4} and Claude \cite{claude} has further pushed the boundaries of code generation capabilities, allowing for more contextually appropriate and functionally correct code outputs. More recently, reasoning-enhanced models have made substantial progress in the code generation domain, with claude-3.7 \cite{claude}, deepSeek-r1 \cite{deepseek-r1}, and o4-mini \cite{o4-mini} exhibiting extraordinary capabilities in producing complex, functional code.

\textbf{Code Generation Benchmarks.} LLM code generation evaluation has evolved through diverse benchmarks. HumanEval \cite{humaneval} established a standard with Python function completion tasks, while MBPP \cite{mbpp} expanded this approach with varied difficulty problems. Specialized benchmarks emerged targeting different programming aspects: CodeContests \cite{codecontest} and APPS \cite{apps} focus on competitive programming challenges, while DS-1000 \cite{ds-1000} addresses data science tasks. Recent developments include LiveCodeBench~\cite{livecodebench} for measuring performance on coding competitions across different time periods, Aider~\cite{aider} evaluate LLM’s ability to follow instructions and edit code successfully without human intervention, and SWE-Bench \cite{swe-bench} for real-world GitHub issues. Open-R1~\cite{openr1} maintained a leaderboard of IOI 2024, but only a few problems were prone to causing errors and overfitting. As illustrated in Figure~\ref{tab:comparebaseline}, \name distinguishes itself through its exceptional difficulty level: even the most advanced reasoning-focused LLMs struggle to perform well, thereby establishing a new ceiling for evaluating the code generation capabilities of current models.

%% file: tables/dataset_comparison.tex
\begin{table}[t]
  \centering
  
  \resizebox{\linewidth}{!}{%
    \begin{tabular}{l@{\hskip 1.5pt}|cc@{\hskip 4.5pt}c@{\hskip 2.5pt}c@{\hskip -4.5pt}c@{\hskip 0pt}}
    \toprule
    \multirow{2}{*}{\textbf{Benchmark}}  & \textbf{Diff'} & \textbf{Lang-} & \textbf{Test} & \textbf{Problem} & \textbf{Human} \\
    & \textbf{level} & \textbf{uage} & \textbf{Cases} & \textbf{Category} & \textbf{Results} \\
    \midrule
    
    OlympicArena &   \progressbar{0.8}{0.8cm}{0.25cm}{teal}  &  -  &  {\color{red}\XSolidBrush} & Interactive & {\color{red}\XSolidBrush} \\
    CodeContest &   \progressbar{0.8}{0.8cm}{0.25cm}{teal}    &   C++        & {\color{green!60!black}\CheckmarkBold}  & Standard I/O &  {\color{red}\XSolidBrush}\\
    LiveCodeBench &   \progressbar{0.6}{0.8cm}{0.25cm}{teal}    &    Python            & {\color{green!60!black}\CheckmarkBold}       & Standard I/O & {\color{red}\XSolidBrush} \\
    
    APPS &   \progressbar{0.4}{0.8cm}{0.25cm}{teal}    & Python        & {\color{green!60!black}\CheckmarkBold}         & Standard I/O & {\color{red}\XSolidBrush} \\

    \midrule
    \multirow{2}{*}{\textbf{HLCE (Ours)}}  &  \multirow{2}{*}{\progressbar{1}{0.8cm}{0.25cm}{teal}}   &  C++,     &   \multirow{2}{*}{{\color{green!60!black}\CheckmarkBold}}   &  Standard I/O & \multirow{2}{*}{{\color{green!60!black}\CheckmarkBold}} \\
    & & Python & & \&Interactive & \\
    \bottomrule
    \end{tabular}%
    }
  
  \caption{HLCE vs other code competition benchmarks.}
  \label{tab:comparebaseline}
\end{table}

%% file: tex/3benchmark.tex
\begin{figure*}[ht]
    \centering
    \includegraphics{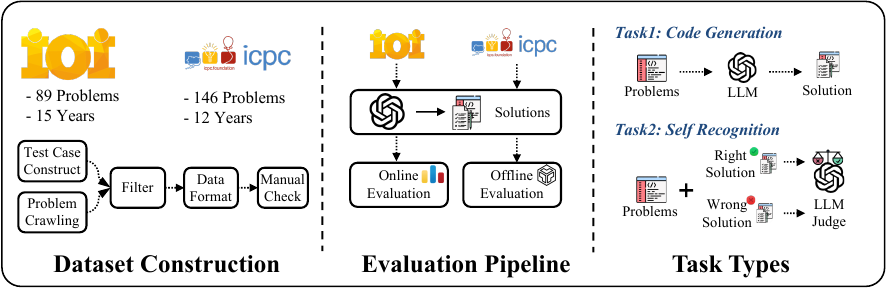}
    \vspace{-1em}
    \caption{\name Benchmark Overview.}
    \label{fig:Overview}
\end{figure*}

\section{\name Benchmark}
As illustrated in Figure~\ref{fig:Overview}, the \name framework comprises three principal components: \textbf{Dataset}, \textbf{Evaluation Task}, and \textbf{Evaluation Framework}.


\subsection{Dataset}
We collect our dataset from two premier competitive programming contests: the International Olympiad in Informatics (IOI) and the International Collegiate Programming Contest (ICPC) World Finals. The competitions represent state-of-the-art algorithmic problem-solving challenges, attracting elite participants worldwide and featuring meticulously designed problems that evaluate advanced computational thinking capabilities.

For \textbf{ICPC World Finals}, our dataset construction process is as follows:
\begin{itemize}[leftmargin=*, itemsep=-0.5em, topsep=0.0em]
    \item \textbf{Extracting Problem Statements}: Since official problem descriptions are only available in PDF format, we first attempted to use various PDF parsing tools (e.g., Markitdown~\cite{markitdown}), but none yielded satisfactory results. Hence, we manually copied and pasted problem content from the PDFs, then used ChatGPT to further refine and standardize the formatting.
    \item \textbf{Collecting Test Cases}: We carefully gathered all official test cases for each problem to ensure completeness and accuracy for evaluation.
    \item \textbf{Data Filtering}: We removed problems with corrupted or missing official test cases (such as some problems from 2018), as well as interactive problems incompatible with the standard ACM input/output format.
\end{itemize}
This rigorous process resulted in a dataset of 146 problems from the 2011--2023 ICPC World Finals, each accompanied by complete test cases.


For \textbf{IOI}, the construction process is as follows:
\begin{itemize}[leftmargin=*, itemsep=-0.5em, topsep=0.0em]
    \item \textbf{Extracting Problem Statements}: We identified a comprehensive dump\footnote{https://ioi.contest.codeforces.com/group/32KGsXgiKA/blog} of IOI problems on Codeforces and systematically extracted all problem descriptions using web scraping techniques. The raw extracted content was cleaned and reformatted using ChatGPT to resolve formatting and consistency issues.
    \item \textbf{Collecting Test Cases}: We gathered all official test cases corresponding to each problem for evaluation purposes.
    \item \textbf{Data Filtering}: Output-only problems, which require only output files instead of program code, were excluded due to evaluation challenges.
\end{itemize}
The resulting IOI dataset includes 89 problems from 2010–2024, each with full test cases and standardized formatting.


\subsection{Evaluation Tasks}
We define two distinct evaluation tasks in our benchmark. The first is the classic \textbf{code generation} task, which serves as a standard metric to evaluate the ability of LLMs to generate accurate code. In addition to this, we propose a novel task termed the \textbf{self-recognition} task, which aims to evaluate the model's ability to recognize whether the code it generates is correct or incorrect. This task provides a unique perspective on the model's reasoning and introspection capabilities, which are crucial for applications requiring reliable and autonomous coding solutions.

\subsection{Evaluation Framework}
For the code generation task on ICPC World Finals problems, we implemented a Python-based evaluation framework that utilizes standard test cases with predefined input/output pairs.

For the IOI dataset, despite having collected all test cases, most IOI problems require interaction between the evaluation system and C++ functions in the submission, necessitating the implementation of specific evaluation programs for each problem. To address this issue, we developed an automated submission bot that interfaces with the Codeforces IOI platform. This bot submits LLM-generated solutions to the official Codeforces IOI judging system and retrieves detailed scoring information and execution results for each problem.

For the self-recognition task, we implemented a separate Python evaluation framework designed to assess LLMs' ability to accurately recognize their own capabilities. 

%% file: tex/4experiments.tex
\section{Experiments}
In this section, we evaluate and analyze the performance of current state-of-the-art LLMs on two tasks from the HLCE benchmark.

\subsection{Experimental Setup}

\textbf{Benchmarked Models.} To conduct a comprehensive evaluation, we assessed a diverse set of SOTA LLMs, encompassing both reasoning and non-reasoning models. For non-reasoning LLMs, we selected \textbf{gpt-4o-mini}, \textbf{claude-3-5-sonnet-20241022}, \textbf{claude-3-7-sonnet-20250219}, \textbf{gpt-4o-2024-05-13}, \textbf{deepseek-v3-0324}, and \textbf{chatgpt-4o-latest}. The reasoning models included \textbf{o1-mini}, \textbf{o1}, \textbf{o3-mini (High)}, \textbf{Gemini-2.5-Pro}, \textbf{DeepSeek-R1}, and \textbf{claude-3.7-sonnet-thinking}.

\textbf{Implementation Details.} For all models, we utilized the OpenRouter\footnote{https://openrouter.ai/} API with default hyperparameters. We generated 5 responses per question and evaluated performance using pass@1~\cite{humaneval} and pass@5 metrics. For IOI questions, scoring is based on test case success, with 100 indicating all tests passed. Scores below 100 are considered failures. Additional details are provided in Appendix~\ref{sec:appendix}.

\subsection{Evaluation Results on \name}

\input{tables/results_main}

\subsubsection{Results on Code Generation Task} 
Based on the results in Table ~\ref{tab:generation}, we can observe several significant trends in the performance of LLMs on the \name code generation task:

\paragraph{Task Difficulty} The results highlight the extraordinary difficulty of our \name benchmark compared to existing code generation tasks. While o4-mini(high) achieved impressive scores on standard benchmarks (98.5\% pass@1 on HumanEval, 74.9\% pass@1 on LivecodeBench, 68.1\% pass@1 on SWE-bench Verified, and 72.0\% on Aider), it only achieved a pass@1 rate of 15.85\% on \name. This significant performance gap reveals several issues: \name problems are exceptionally challenging, particularly those from IOI (6.48\% pass@1 rate), requiring advanced computational thinking that current Large Language Models (LLMs) struggle to master; the gap between pass@1 and pass@5 rates (15.85\% vs. 29.31\%) indicates that models possess the necessary knowledge but lack reasoning consistency. This difficulty establishes \name as a critical benchmark for measuring advanced reasoning and algorithmic problem-solving capabilities in LLMs.


\paragraph{Non-reasoning vs. Reasoning Models} A significant performance gap exists between non-reasoning and reasoning-enhanced LLMs. Almost all reasoning models consistently outperform non-reasoning counterparts across all metrics and datasets. The best reasoning model (o4-mini(high)) achieves 15.85\% average pass@1, approximately 4.5 times higher than the best non-reasoning model (deepseek-v3-0324) at 3.53\%. Notably, deepseek-v3-0324 exhibits superior performance among non-reasoning models, even surpassing the reasoning model Claude-3.7-Thinking. We attribute this capability to DeepSeek-V3-0324 being distilled from DeepSeek-R1 data, which enhanced its coding abilities. This suggests that targeted distillation from reasoning-rich data could yield models with advanced coding capabilities without requiring explicit reasoning during inference.

\paragraph{Model degradation Phenomenon}
An interesting exception to the general trend is the claude-3.7-thinking model, which underperforms compared to non-reasoning models and shows notably weaker results on IOI problems than both claude-3.5-sonnet and claude-3.7-sonnet. This model achieves 0\% pass rates on IOI problems, representing a significant decline despite its enhanced reasoning capabilities. We hypothesize this degradation stems from Anthropic's optimization focus on general software engineering rather than competitive programming. The claude-3.7 technical report~\cite{claude} indicates that Claude 3.7 Sonnet achieves 62.3\% accuracy on SWE-bench Verified, outperforming o3-mini(high) (49.3\%), suggesting a deliberate trade-off favoring practical software engineering tasks.
\paragraph{Standard I/O vs. Interactive Evaluation} 
A striking observation is the significant performance gap of models between IOI and ICPC World Finals competitions. Even o4-mini(high), which achieves a pass@1 rate of 25.21\% on ICPC World Finals, only manages 6.48\% on IOI. We hypothesize that this substantial discrepancy stems from the training methodology of current reasoning models, which predominantly utilize Standard I/O-based data with outcome-reward RL. This approach aligns well with most ICPC problems but fails to address the interactive nature of IOI problems, which often require program interaction with a judge. This finding underscores the importance of incorporating more diverse training data and developing more robust RL training environments to enhance model performance across different problem types.

\subsubsection{Results on Self-recognition Task}  The self-recognition task evaluates models' ability to accurately judge the correctness of their own solutions, providing insights into their metacognitive capabilities. Table~\ref{tab:self-recognition} presents the AUC scores for this task across various models.

\input{tables/results_recognition}

\paragraph{Performance Comparison}
Among non-reasoning models, ChatGPT-4o-latest demonstrates superior self-recognition with an AUC of 0.84, despite showing only moderate performance in code generation (pass@1: 1.18\%, pass@5: 2.37\%). This suggests that strong self-recognition capability does not necessarily correlate with superior code generation performance. Similarly, DeepSeek-R1 achieves the highest AUC (0.81) among reasoning models while ranking in the middle tier for code generation.

\paragraph{Knowledge of Self-Knowledge: The Socratic Paradox in LLMs}
The Socratic paradox, encapsulated in the statement "I know that I know nothing," represents the philosophical understanding that true wisdom begins with recognizing the limits of one's knowledge. However, across the two tasks in \name, we observe a contrasting phenomenon. Interestingly, the top-performing model in self-recognition comes from the non-reasoning category (\textbf{chatgpt-4o-latest: 0.84}), outperforming all reasoning models. This contrasts with the code generation task, where reasoning models (particularly o4-mini) significantly outperformed non-reasoning counterparts. This empirical evidence reveals a fundamental challeng: some models excel at problem-solving but lack accurate self-recognition, while others show better self-recognition despite lower performance. This disconnect suggests reasoning abilities and self-recognition develop along different trajectories in current LLM architectures, highlighting the need for research that enhances both dimensions simultaneously.



\subsection{Test Time Scaling Law on \name}
\begin{figure*}[htbp]
    \centering
    \begin{subfigure}{0.95\linewidth}
        \caption{Pass@1 vs. Average Output Tokens in Non-reasoning Models}
        \includegraphics[width=\linewidth]
        {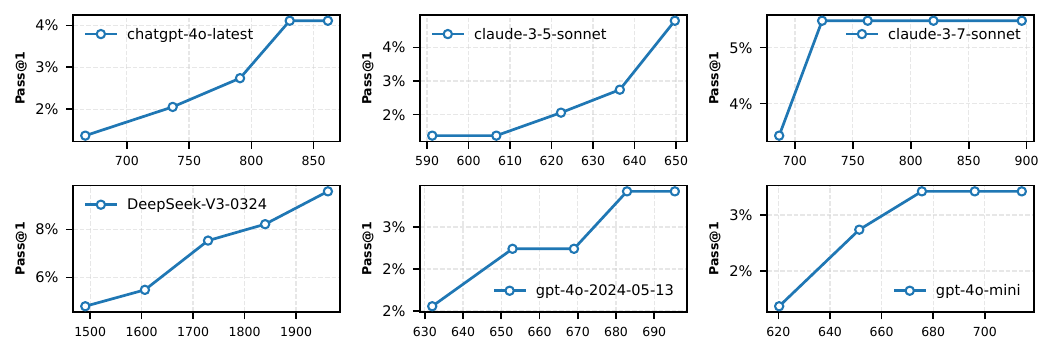}
        \label{fig:scaling_law_non_reasoning}
        \vspace{-1em}
    \end{subfigure}
    \begin{subfigure}{0.95\linewidth}
        \centering
        \vspace{-1em}
        \caption{Pass@1 vs. Average Output Tokens in Reasoning Models}
        \includegraphics[width=\linewidth]
        {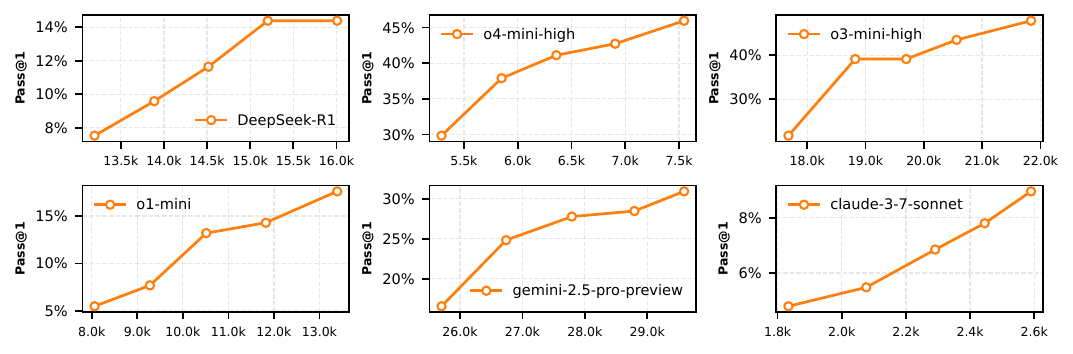}
        \label{fig:scaling_law_reasoning}
    \end{subfigure}
    \vspace{-2em}
    \caption{Comparison of Non-Reasoning and Reasoning Models in Test-time Scaling.}
    \label{fig:scaling_law}
\end{figure*}

Test time scaling law, where increased computational resources during inference improve model performance \cite{o1,openr1,o4-mini}, has been validated on Olympic-level mathematics \cite{s1} but rarely on Olympic-level programming challenges, except by OpenAI \cite{openaitest}.
 Therefore, in this section, we utilize the extremely difficult \name benchmark to verify that the test time scaling law still holds for current SOTA LLMs.  
Specifically, we group all generated responses by token count and calculate the pass@1 rate for each group in ICPC World Finals. The results are presented in Figures~\ref{fig:scaling_law_non_reasoning} and ~\ref{fig:scaling_law_reasoning}. 

\paragraph{Test Time Scaling Law Holds} As shown in the figures, both models with and without reasoning capabilities demonstrate a clear test time scaling law on the \name benchmark. Pass@1 rate gradually increases as thinking time increases. This indicates that complex programming problems benefit significantly from extended reasoning processes, and models can leverage additional computation time to refine their solutions and correct potential errors.
\paragraph{Test Time Scaling Law has not yet reached its Boundary} Even for the most capable models such as o4-mini(high) and Gemini-2.5-pro, the pass@1 rate continues to increase with longer reasoning sequences. This suggests that despite the impressive capabilities of current state-of-the-art models, the test time scaling law \textbf{has not yet reached its limit}. Surprisingly, we observed that compared to o3-mini(high), o4-mini(high) reduced the average output token length by approximately two-thirds. This finding has important implications for the future development of code LLMs, indicating that we can confidently continue to optimize test-time scaling laws to further enhance model performance.

\subsection{Can LLMs Surpass Top-tier Human?}
To assess whether the latest LLMs can truly compete with elite human programmers, we compared the performance of SOTA LLMs against the gold, silver, and bronze medal thresholds from historical IOI and ICPC World Finals competitions.  The comparative results are presented in Tables~\ref{tab:ioi_performance} and~\ref{tab:icpc_performance}.


\paragraph{LLMs Could Reach  Medal Level} From Tables~\ref{tab:ioi_performance} and~\ref{tab:icpc_performance}, we observe a remarkable finding: current sota models can achieve medal-winning performance in prestigious programming competitions such as IOI and ICPC World Finals. Notably, Gemini-2.5-pro and o4-mini(high) demonstrate exceptional capabilities, securing silver and gold medals in IOI and ICPC World Finals, respectively. This indicates that contemporary SOTA LLMs have developed computational reasoning capabilities that rival those of top human competitors. Furthermore, Gemini-2.5-Pro exhibits impressive consistency in IOI performance, earning five bronze medals and one silver medal. These results suggest that in the near future, we may reasonably anticipate LLMs achieving gold medal performance levels across such competitions.

\input{tables/results_ioi}

\paragraph{Discrepancy Between Medal Achievement and Pass@1 Rate} The intriguing discrepancy observed in Table~\ref{tab:ioi_performance} and Table~\ref{tab:generation}, where models like o4-mini and Gemini-2.5-Pro can achieve medal-worthy scores in IOI competitions while maintaining relatively low pass@1 rates, highlights a critical phenomenon in LLM code generation capabilities. We select the highest score from five submissions as the model's total score, indicating that these models possess the knowledge required to solve problems correctly, yet struggle to produce correct solutions in a single attempt, resulting in low pass@1 rates. This finding underscores the importance of developing more reliable methods to guide LLMs' problem-solving capabilities as a crucial direction for future research.

\input{tables/results_icpc}


\subsection{Efforts to Prevent Data Leakage}
The critical issue of LLMs overfitting~\cite{livecodebench}to benchmark leaderboards remains inadequately addressed. To ensure fair evaluation, we established a private leaderboard using 2024 ICPC World Finals problems with unreleased test cases, for which we constructed our own test cases, creating an unbiased assessment environment.

Our evaluation employs a two-tier system: models submitted to the public leaderboard undergo automatic testing on our private dataset. This approach reveals potential overfitting by comparing cross-leaderboard performance. Models with consistent results likely demonstrate authentic code generation capabilities, while performance disparities may indicate benchmark-specific overfitting. 

%% file: tables/results_main.tex
\begin{table*}[htb]
\centering
 \label{tab:model_performance_comparison}
\renewcommand{\arraystretch}{1.0
} 
\begin{tabular}{l|cc|cc|cc}
  \toprule
\multirow{2}{*}{\textbf{Model}} &\multicolumn{2}{c|}{\textbf{ICPC World Finals}} & \multicolumn{2}{c|}{\textbf{IOI}} & \multicolumn{2}{c}{\textbf{Avg.}}\\
     & pass@1 & pass@5 & pass@1 & pass@5 & pass@1 & pass@5 \\
  \midrule
  \rowcolor[HTML]{F3F3F3}\multicolumn{7}{c}{\itshape Non-reasoning}\\
  \midrule 
 \textbf{gpt-4o-mini}               & 0.96 & 2.99 & 0.00 & 0.00 & 0.48 & 1.50 \\ 
 \textbf{claude-3.5-sonnet}         & 2.74 & 5.04 & 0.67 & 1.12 & 1.71 & 3.08 \\ 
 \textbf{claude-3.7-sonnet}         & 3.84 & 6.41 & \textbf{1.12} & 1.12 & 2.48 & 3.77 \\ 
 \textbf{gpt-4o-2024-05-13}         & 1.99 & 3.35 & 0.45 & 1.12 & 1.22 & 2.24 \\ 
 \textbf{deepseek-v3-0324}          & \textbf{6.16} & \textbf{12.1} & 0.90 & 1.12 & \textbf{3.53} & \textbf{6.61} \\ 
 \textbf{chatgpt-4o-latest}         & 1.91 & 3.61 & 0.45 & 1.12 & 1.18 & 2.37 \\ 
  \midrule
  \rowcolor[HTML]{F3F3F3}\multicolumn{7}{c}{\itshape Reasoning}\\
  \midrule
  \textbf{claude-3.7-thinking}      & 4.25 & 8.22 & 0.00 & 0.00 & 2.13 & 4.11 \\  
 \textbf{o1-mini}                   & 10.55 & 19.86 & 2.34 & 3.37 & 6.45 & 11.62 \\ 
 \textbf{DeepSeek-R1}               & 8.08 & 14.38 & 2.23 & 5.62 & 5.16 & 10.00 \\ 
 \textbf{o3-mini (High)}            & 13.42 & 29.45 & \textbf{8.26} & 10.23 & 10.84 & 19.84 \\ 
 \textbf{Gemini-2.5-Pro}            & 17.40 & 29.45 & 5.39 & 11.24 & 11.40 & 20.35 \\ 
 \textbf{o4-mini (High)}            & \textbf{25.21} & \textbf{43.84} & 6.48 & \textbf{14.77} & \textbf{15.85} & \textbf{29.31} \\ 
 \bottomrule
\end{tabular}
\vspace{-0.5em}
\caption{Performance results on \name code generation task.}
\label{tab:generation}
\end{table*}

%% file: tables/results_recognition.tex
\begin{table}[htb]
\centering
\resizebox{0.75\linewidth}{!}{ 
\begin{tabular}{ll|c}
\toprule
\multicolumn{2}{c|}{\textbf{Model Name}} & \textbf{AUC$\uparrow$} \\ 
\midrule

\parbox[t]{3.0mm}{\multirow{6}{*}{\rotatebox[origin=c]{90}{\textbf{Non-reasoning}}}}

& {gpt-4o-mini}         &  0.60   \\
& {claude-3.5-sonnet}   &  0.75    \\
& {claude-3.7-sonnet}   &   0.69    \\
& {gpt-4o-2024-05-13}   &  0.76     \\ 
& {deepseek-v3-0324}    &  0.63  \\
& \textbf{chatgpt-4o-latest}   &   \textbf{0.84}  \\ 
\midrule

\parbox[t]{3.0mm}{\multirow{6}{*}{\rotatebox[origin=c]{90}{\textbf{Reasoning}}}}
& {claude-3.7-thinking} & 0.79  \\
& {o1-mini}             &  0.73  \\
& \textbf{DeepSeek-R1}  &  \textbf{0.81}  \\
& {o3-mini (High)}      &  0.66 \\
& {Gemini-2.5-Pro}      &  0.72   \\
& {o4-mini (High)}      &  0.63 \\

\bottomrule

\end{tabular}
}
\vspace{-0.5em}
\caption{Performance results on self-recognition task, $\uparrow$ denotes higher is better.}
\label{tab:self-recognition}
\vspace{-1.5em}
\end{table}

%% file: tables/results_ioi.tex
\begin{table}[htbp]
\small
\setlength{\tabcolsep}{1.5pt}

\resizebox{\linewidth}{!}{%
\begin{tabular}{c|ccc|ccc}
\toprule
\multicolumn{7}{c}{IOI(Max Points)}                                                                 \\ \midrule
\multirow{2}{*}{Year} &  Bronze & Silver  & Gold & o3-mini & Gemini  & o4-mini \\ 
 &  {\color{brown}\faMedal} & {\color{gray}\faMedal} & {\color{yellow}\faMedal} & (high) & 2.5 pro & (high) \\
\midrule

2024 & 217 & 289 & 360 & 133 & 138 & 89  \\
2023 & 153 & 230 & 334 & 53  & 157{\color{brown}\faMedal} & 52 \\
2022 & 147 & 258 & 416 & 5   & 150{\color{brown}\faMedal} & 138 \\
2021 & 203 & 289 & 373 & 220{\color{brown}\faMedal} &  292{\color{gray}\faMedal} & 219{\color{brown}\faMedal} \\
2020 & 236 & 338 & 480 & 92  & 237{\color{brown}\faMedal} & 119 \\
2016 & 240 & 328 & 416 & 212 & 317{\color{brown}\faMedal} & 12  \\
2015 & 185 & 326 & 440 & 211{\color{brown}\faMedal} & 119 & 162 \\
2014 & 223 & 323 & 449 & 38  & 224{\color{brown}\faMedal} & 150 \\
2013 & 220 & 359 & 480 & 100 & 203 & 323{\color{brown}\faMedal}  \\
2012 & 157 & 237 & 364 & 40  & 224 & 200{\color{brown}\faMedal}  \\
2011 & 267 & 370 & 478 & 0   & 224 & 0 \\
\bottomrule
\end{tabular}}
\caption{Human Performance thresholds for medals and the best LLM Performance in IOI. We utilize the highest score from the five responses.}
\vspace{-1em}
\label{tab:ioi_performance}
\end{table}

%% file: tables/results_icpc.tex
\begin{table}[htbp]
\small
\setlength{\tabcolsep}{1.5pt}
\resizebox{\linewidth}{!}{%
\begin{tabular}{c|ccc|ccc}
\toprule
\multicolumn{7}{c}{ICPC World Finals(Solved Problems)}                                                                 \\ \midrule
\multirow{2}{*}{Year} &  Bronze & Silver  & Gold & o3-mini & Gemini  & o4-mini \\ 
 &  {\color{brown}\faMedal} & {\color{gray}\faMedal} & {\color{yellow}\faMedal} & (high) & 2.5 pro & (high) \\
\midrule

2023 & 7 & 8 & 9 & 3 & 3 & 3   \\
2022 & 8 & 8 & 9 & 2 & 3 & 4   \\
2021 & 8 & 8 & 9 & 5 & 4 & 5   \\
2020 & 9 & 10& 11& 2 & 4 & 6   \\
2019 & 7 & 7 & 8 & 1 & 2 & 2   \\
2018 & 7 & 7 & 8 & 0 & 1 & 2   \\
2017 & 8 & 8 & 10& 4 & 3 & 4   \\
2016 & 9 & 9 & 10& 5 & 4 & 6   \\
2015 & 9 & 10& 10& 9 {\color{brown}\faMedal} & 7 & 8 \\
2014 & 4 & 5 & 6 & 2 & 4 & 4  \\
2013 & 6 & 7 & 8 & 3 & 1 & 6  \\
2012 & 6 & 7 & 7 & 4 & 3 & 8 {\color{yellow}\faMedal}  \\
2011 & 7 & 7 & 8 & 3 & 4 & 6  \\
\bottomrule
\end{tabular}}
\caption{Human Performance thresholds for medals and the best LLM Performance in ICPC World Finals. A problem is considered solved if any of the five responses solves it.}
\label{tab:icpc_performance}
\vspace{-2em}
\end{table}

%% file: tex/5dicussion.tex
\section{Discussion}
\paragraph{The Future Direction of Code LLM Capabilities} The code LLM landscape is rapidly evolving with divergent approaches to balancing general coding capabilities and competitive programming skills. Claude 3.7 appears to prioritize general code generation over specialized algorithmic performance, possibly reflecting Anthropic's assessment of market demand for practical development assistance. Conversely, Gemini 2.5 Pro excels in both domains, suggesting that with sufficient scale and sophistication, the trade-off between these capabilities may be less constraining than previously thought.  Therefore, general code abilities and competitive programming abilities in future LLMs are \textbf{not mutually exclusive directions}. LLM performance in programming competitions remains an important metric for gauging current progress toward AGI.

\paragraph{The Future Direction of Test-Time Scaling in Code LLM} With the powerful performance demonstrated by Gemini 2.5-Pro and o4-mini (high), concerns have emerged about whether Test-time Scaling Laws, similar to Pre-training Scaling Laws~\cite{scalinglaw}, are approaching their limits. Our experiments demonstrate that test-time scaling substantially impacts model performance. While these results are impressive, we have likely not reached the ceiling of potential improvements through such techniques. As models continue to scale in test time, we anticipate further advancements in reasoning efficiency~\cite{dumitru2025copyspec} and efficacy, and generating sophisticated algorithmic solutions.

%% file: figures/icpc_question_response.tex
\begin{figure*}[ht]
\begin{tcolorbox}[
    title=Example of ICPC World Finals problems,
    colframe=gray!60!black,
    colback=white,
    boxrule=0.5pt,
    sharp corners,
    left=3mm, right=3mm, top=2mm, bottom=2mm,
    fonttitle=\bfseries
]
\small
\textbf{QuestionId:2021-H}

\textbf{QuestionName:Mining Your Own Business}

\textbf{Problem Description}

John Digger is the owner of a large illudium phosdex mine. The mine is made up of a series of tunnels that meet at various large junctions. Unlike some owners, Digger actually cares about the welfare of his workers and has a concern about the layout of the mine. Specifically, he worries that there may be a junction which, in case of collapse, will cut off workers in one section of the mine from other workers (illudium phosdex, as you know, is highly unstable). To counter this, he wants to install special escape shafts from the junctions to the surface. He could install one escape shaft at each junction, but Digger doesn't care about his workers that much. Instead, he wants to install the minimum number of escape shafts so that if any of the junctions collapses, all the workers who survive the junction collapse will have a path to the surface.

Write a program to calculate the minimum number of escape shafts and the total number of ways in which this minimum number of escape shafts can be installed.

\textbf{Input}

The input consists of several test cases. The first line of each case contains a positive integer \( N \) (\( N \leq 5 \cdot 10^4 \)) indicating the number of mine tunnels. Following this are \( N \) lines each containing two distinct integers \( s \) and \( t \), where \( s \) and \( t \) are junction numbers. Junctions are numbered consecutively starting at 1. Each pair of junctions is joined by at most a single tunnel. Each set of mine tunnels forms one connected unit (that is, you can get from any one junction to any other).

The last test case is followed by a line containing a single zero.

\textbf{Output}

For each test case, display its case number followed by the minimum number of escape shafts needed for the system of mine tunnels and the total number of ways these escape shafts can be installed. You may assume that the result fits in a signed 64-bit integer.

Follow the format of the sample output.

\textbf{Sample Input}
\begin{verbatim}
9
1 3
4 1
3 5
1 2
2 6
1 5
6 3
1 6
3 2
6
1 2
1 3
2 4
2 5
3 6
3 7
0
\end{verbatim}

\textbf{Sample Output}
\begin{verbatim}
Case 1: 2 4
Case 2: 4 1
\end{verbatim}
\end{tcolorbox}
\vspace{-1em}
\caption{Example of ICPC World Finals problem.}
\label{fig:icpc_problem}
\end{figure*}

%% file: figures/IOI_problems.tex
\begin{figure*}[t]
\begin{tcolorbox}[
    title=Example of IOI Problem,
    colframe=gray!60!black,
    colback=white,
    boxrule=0.5pt,
    sharp corners,
    left=3mm, right=3mm, top=2mm, bottom=2mm,
    fonttitle=\bfseries
]
\small 
\textbf{ProblemID:2024 D. Hieroglyphs}

\textbf{Time limit:} 2 seconds \\
\textbf{Memory limit:} 1024 megabytes \\
\textbf{Input/Output:} standard

\textbf{Problem Statement:}

A team of researchers is studying the similarities between sequences of hieroglyphs. They represent each hieroglyph with a non-negative integer. To perform their study, they use the following concepts about sequences.

For a fixed sequence $A$, a sequence $S$ is called a \textit{subsequence} of $A$ if and only if $S$ can be obtained by removing some elements (possibly none) from $A$.

The table below shows some examples of subsequences of a sequence $A = [3, 2, 1, 2]$.
\begin{center}
\begin{tabular}{|l|l|}
\hline
Subsequence & How it can be obtained from $A$ \\
\hline
$[3, 2, 1, 2]$ & No elements are removed. \\
$[2, 1, 2]$ & [$\underline{3}$, 2, 1, 2] \\
$[3, 2, 2]$ & [3, 2, $\underline{1}$, 2] \\
$[3, 2]$ & [3, $\underline{2}$, $\underline{1}$, 2] or [3, 2, $\underline{1}$, $\underline{2}$] \\
$[3]$ & [3, $\underline{2}$, $\underline{1}$, $\underline{2}$] \\
$[ ]$ & [$\underline{3}$, $\underline{2}$, $\underline{1}$, $\underline{2}$] \\
\hline
\end{tabular}
\end{center}

...

\textbf{Implementation details:}

You should implement the following procedure:
\begin{verbatim}
std::vector<int> ucs(std::vector<int> A, std::vector<int> B)
\end{verbatim}

\begin{itemize}
\item $A$: array of length $N$ describing the first sequence.
\item $B$: array of length $M$ describing the second sequence.
\item If there exists a universal common subsequence of $A$ and $B$, the procedure should return an array containing this sequence. Otherwise, the procedure should return $[-1]$ (an array of length $1$, whose only element is $-1$).
\end{itemize}

This procedure is called exactly once for each test case.

\textbf{Input:}

The sample grader reads in the following format:
\begin{itemize}
\item line $1$: $N$ $M$ ($1 \leq N \leq 100\,000$, $1 \leq M \leq 100\,000$)
\item line $2$: $A[0]\; A[1]\ldots A[N-1]$ ($0 \leq A[i] \leq 200\,000$)
\item line $3$: $B[0]\; B[1]\ldots B[M-1]$ ($0 \leq B[j] \leq 200\,000$)
\end{itemize}

\textbf{Output:}

The sample grader prints in the following format:
\begin{itemize}
\item line $1$: $T$
\item line $2$: $R[0]\; R[1]\ldots R[T-1]$
\end{itemize}

Here, $R$ is the array returned by \texttt{ucs} and $T$ is its length.

\textbf{Sample Input:}
\begin{verbatim}
6 5
0 0 1 0 1 2
2 0 1 0 2
\end{verbatim}

\textbf{Sample Output:}
\begin{verbatim}
4
0 1 0 2
\end{verbatim}
\end{tcolorbox}
\vspace{-0.5em}
\caption{Example of IOI  problem.}
\label{fig:ioi_problem}
\end{figure*}

%% file: tables/api_cost.tex
\begin{table}[h]
\centering
\begin{tabular}{lr}
\hline
\textbf{Model} & \textbf{Cost (USD)} \\
\hline
Gemini-2.5-Pro & 758.11 \\
DeepSeek-R1 & 359.55 \\
o3-mini & 239.70 \\
claude-3.7-sonnet-thinking & 118.44 \\
o1-mini & 109.86 \\
o4-mini & 77.90 \\
claude-3-7-sonnet & 39.48 \\
gpt-4o-2024-05-13 & 25.67 \\
claude-3-5-sonnet & 23.37 \\
chatgpt-4o-latest & 19.98 \\
deepseek-v3 & 16.98 \\
gpt-4o-mini & 1.02 \\
\hline
\textbf{Total} & \textbf{1,790.06} \\
\hline
\end{tabular}
\caption{API costs for our experiments.}
\label{tab:api_cost}
\end{table}

%% file: figures/IOI_prompt.tex
\begin{figure*}[ht]
\begin{AIbox}{Prompt for IOI code generation}

\begin{tcolorbox}[
    colframe=gray!60!black,
    colback=white,
    boxrule=0.4pt,
    sharp corners,
    left=1mm, right=1mm, top=0.5mm, bottom=0.5mm,
]
\small\ttfamily
You are now an expert contestant in the International Olympiad in Informatics (IOI). For most problems, please implement a C++ solution for the given problem with the following guidelines: 

- You will be given a problem statement, test case constraints and example test inputs and outputs. Please reason step by step about the solution, then provide a complete implementation in C++.

- You should correctly implement the routine(s) described in Implementation Details, without reading or writing anything directly from stdin or to stdout, as input and output are passed through the implemented routines.

- Assume your code will be run on the OFFICIAL grader, and do not add a main, a sample grader, or any other test function unless it has been explicitly requested.

- IMPORTANT: When implementing functions required by the problem description that use notation like \texttt{int[]} or \texttt{int64[]} for function parameters, implement them as C++ std::vector types (e.g., \texttt{vector<int>} or \texttt{vector<long long>}), not as raw arrays or pointers.

- When declaring or implementing functions that are provided by the grader, use the EXACT same parameter types as specified. Do not use const references (\texttt{const vector<int>\&}) or non-const references (\texttt{vector<int>\&}) when the grader expects \texttt{vector<int>}, even if it would be more efficient.

- For multi-dimensional arrays like \texttt{int[][]}, implement them as nested vectors (e.g., \texttt{vector<vector<int>>}) without references.

Please place your code between the following delimiters:

\begin{verbatim}
```cpp
// Your code will be placed here
```
\end{verbatim}
\end{tcolorbox}

\end{AIbox}
\vspace{-1em}
\caption{Prompt for IOI code generation tasks.}
\label{fig:pr}
\label{fig:prompt_ioi}
\end{figure*}

%% file: figures/ICPC_world_finals.tex
\begin{figure*}[ht]
\begin{AIbox}{Prompt for ICPC World Finals code generation}

\begin{tcolorbox}[
    colframe=gray!60!black,
    colback=white,
    boxrule=0.4pt,
    sharp corners,
    left=1mm, right=1mm, top=0.5mm, bottom=0.5mm,
]
\small\ttfamily
You are an expert Python programmer.

- You will be given a problem statement, test case constraints and example test inputs and outputs. 

- You will generate a correct Python program that matches the specification and passes all tests

- Read the inputs from stdin solve the problem and write the answer to stdout (do not directly test on the sample inputs). Enclose your code within delimiters as follows. Ensure that when the python program runs, it reads the inputs, runs the algorithm and writes output to STDOUT.

Please place your code between the following delimiters:

\begin{verbatim}
```python
# YOUR CODE HERE
```
\end{verbatim}
\end{tcolorbox}

\end{AIbox}
\vspace{-1em}
\caption{Prompt for ICPC World Finals code generation tasks.}
\label{fig:icpc_generation_prompts}
\end{figure*}

%% file: figures/self-recog.tex
\begin{figure*}[ht]
\begin{AIbox}{Prompt for Self-recognition Tasks}

\begin{tcolorbox}[
    colframe=gray!60!black,
    colback=white,
    boxrule=0.4pt,
    sharp corners,
    left=1mm, right=1mm, top=0.5mm, bottom=0.5mm,
]
\small\ttfamily
You are now a \textbf{code review expert}. I will provide you with a programming problem description along with a corresponding code implementation. Your task is to:

1. Carefully read and understand the problem requirements;\\  
2. Analyze whether the code logic correctly fulfills the problem's requirements;\\  
3. Determine whether the code can pass \textbf{all test cases}, including regular cases, edge cases, and potential hidden tests;\\  
4. In your response, \textbf{first provide a detailed explanation of your analysis}, including any strengths, potential issues, or bugs you identify;\\  
5. Finally, give your conclusion — it must be \textbf{either} of the following two options, and should be wrapped using the delimiter below:

\begin{verbatim}
```answer
Yes
```
\end{verbatim}

or

\begin{verbatim}
```answer
No
```
\end{verbatim}
\end{tcolorbox}

\end{AIbox}
\vspace{-1em}
\caption{Prompt  for Self-recognition tasks.}
\label{fig:prompt_self_reco}
\end{figure*}

%% file: acl_latex.bbl
\begin{thebibliography}{30}
\providecommand{\natexlab}[1]{#1}

\bibitem[{Achiam et~al.(2023)Achiam, Adler, Agarwal, Ahmad, Akkaya, Aleman, Almeida, Altenschmidt, Altman, Anadkat et~al.}]{gpt4}
Josh Achiam, Steven Adler, Sandhini Agarwal, Lama Ahmad, Ilge Akkaya, Florencia~Leoni Aleman, Diogo Almeida, Janko Altenschmidt, Sam Altman, Shyamal Anadkat, and 1 others. 2023.
\newblock Gpt-4 technical report.
\newblock \emph{arXiv preprint arXiv:2303.08774}.

\bibitem[{Aider(2024)}]{aider}
Aider. 2024.
\newblock \href {https://aider.chat/docs/leaderboards/} {Leaderboards - aider documentation}.

\bibitem[{Anthropic(2025)}]{claude}
Anthropic. 2025.
\newblock \href {https://www.anthropic.com/news/claude-3-family} {Introducing the claude 3 model family}.

\bibitem[{Austin et~al.(2021)Austin, Odena, Nye, Bosma, Michalewski, Dohan, Jiang, Cai, Terry, Le et~al.}]{mbpp}
Jacob Austin, Augustus Odena, Maxwell Nye, Maarten Bosma, Henryk Michalewski, David Dohan, Ellen Jiang, Carrie Cai, Michael Terry, Quoc Le, and 1 others. 2021.
\newblock Program synthesis with large language models.
\newblock \emph{arXiv preprint arXiv:2108.07732}.

\bibitem[{Chen et~al.(2021)Chen, Tworek, Jun, Yuan, Pinto, Kaplan, Edwards, Burda, Joseph, Brockman et~al.}]{humaneval}
Mark Chen, Jerry Tworek, Heewoo Jun, Qiming Yuan, Henrique Ponde De~Oliveira Pinto, Jared Kaplan, Harri Edwards, Yuri Burda, Nicholas Joseph, Greg Brockman, and 1 others. 2021.
\newblock Evaluating large language models trained on code.
\newblock \emph{arXiv preprint arXiv:2107.03374}.

\bibitem[{Dong et~al.(2023)Dong, Sun, Kim, and Li}]{dong-etal-2023-open}
Kuicai Dong, Aixin Sun, Jung-jae Kim, and Xiaoli Li. 2023.
\newblock \href {https://doi.org/10.18653/v1/2023.emnlp-main.951} {Open information extraction via chunks}.
\newblock In \emph{Proceedings of the 2023 Conference on Empirical Methods in Natural Language Processing}, pages 15390--15404, Singapore. Association for Computational Linguistics.

\bibitem[{Dumitru et~al.(2025)Dumitru, Yang, Yadav, and Surdeanu}]{dumitru2025copyspec}
Razvan-Gabriel Dumitru, Minglai Yang, Vikas Yadav, and Mihai Surdeanu. 2025.
\newblock \href {https://arxiv.org/abs/2502.08923} {Copyspec: Accelerating llms with speculative copy-and-paste without compromising quality}.
\newblock \emph{Preprint}, arXiv:2502.08923.

\bibitem[{Duong et~al.(2025)Duong, Yang, and Zhang}]{duong2025improving}
Thang Duong, Minglai Yang, and Chicheng Zhang. 2025.
\newblock \href {https://arxiv.org/abs/2505.10861} {Improving the data-efficiency of reinforcement learning by warm-starting with llm}.
\newblock \emph{Preprint}, arXiv:2505.10861.

\bibitem[{El-Kishky et~al.(2025)El-Kishky, Wei, Saraiva, Minaiev, Selsam, Dohan, Song, Lightman, Clavera, Pachocki et~al.}]{openaitest}
Ahmed El-Kishky, Alexander Wei, Andre Saraiva, Borys Minaiev, Daniel Selsam, David Dohan, Francis Song, Hunter Lightman, Ignasi Clavera, Jakub Pachocki, and 1 others. 2025.
\newblock Competitive programming with large reasoning models.
\newblock \emph{arXiv preprint arXiv:2502.06807}.

\bibitem[{Face(2025)}]{openr1}
Hugging Face. 2025.
\newblock \href {https://github.com/huggingface/open-r1} {Open r1: A fully open reproduction of deepseek-r1}.

\bibitem[{Guo et~al.(2025)Guo, Yang, Zhang, Song, Zhang, Xu, Zhu, Ma, Wang, Bi et~al.}]{deepseek-r1}
Daya Guo, Dejian Yang, Haowei Zhang, Junxiao Song, Ruoyu Zhang, Runxin Xu, Qihao Zhu, Shirong Ma, Peiyi Wang, Xiao Bi, and 1 others. 2025.
\newblock Deepseek-r1: Incentivizing reasoning capability in llms via reinforcement learning.
\newblock \emph{arXiv preprint arXiv:2501.12948}.

\bibitem[{Guo et~al.(2024)Guo, Zhu, Yang, Xie, Dong, Zhang, Chen, Bi, Wu, Li et~al.}]{deepseek-coder}
Daya Guo, Qihao Zhu, Dejian Yang, Zhenda Xie, Kai Dong, Wentao Zhang, Guanting Chen, Xiao Bi, Yu~Wu, YK~Li, and 1 others. 2024.
\newblock Deepseek-coder: When the large language model meets programming--the rise of code intelligence.
\newblock \emph{arXiv preprint arXiv:2401.14196}.

\bibitem[{Hendrycks et~al.(2021)Hendrycks, Basart, Kadavath, Mazeika, Arora, Guo, Burns, Puranik, He, Song et~al.}]{apps}
Dan Hendrycks, Steven Basart, Saurav Kadavath, Mantas Mazeika, Akul Arora, Ethan Guo, Collin Burns, Samir Puranik, Horace He, Dawn Song, and 1 others. 2021.
\newblock Measuring coding challenge competence with apps.
\newblock \emph{arXiv preprint arXiv:2105.09938}.

\bibitem[{Hoffmann et~al.(2022)Hoffmann, Borgeaud, Mensch, Buchatskaya, Cai, Rutherford, Casas, Hendricks, Welbl, Clark et~al.}]{scalinglaw}
Jordan Hoffmann, Sebastian Borgeaud, Arthur Mensch, Elena Buchatskaya, Trevor Cai, Eliza Rutherford, Diego de~Las Casas, Lisa~Anne Hendricks, Johannes Welbl, Aidan Clark, and 1 others. 2022.
\newblock Training compute-optimal large language models.
\newblock \emph{arXiv preprint arXiv:2203.15556}.

\bibitem[{Hui et~al.(2024)Hui, Yang, Cui, Yang, Liu, Zhang, Liu, Zhang, Yu, Lu et~al.}]{qwen2.5coder}
Binyuan Hui, Jian Yang, Zeyu Cui, Jiaxi Yang, Dayiheng Liu, Lei Zhang, Tianyu Liu, Jiajun Zhang, Bowen Yu, Keming Lu, and 1 others. 2024.
\newblock Qwen2. 5-coder technical report.
\newblock \emph{arXiv preprint arXiv:2409.12186}.

\bibitem[{Jaech et~al.(2024)Jaech, Kalai, Lerer, Richardson, El-Kishky, Low, Helyar, Madry, Beutel, Carney et~al.}]{o1}
Aaron Jaech, Adam Kalai, Adam Lerer, Adam Richardson, Ahmed El-Kishky, Aiden Low, Alec Helyar, Aleksander Madry, Alex Beutel, Alex Carney, and 1 others. 2024.
\newblock Openai o1 system card.
\newblock \emph{arXiv preprint arXiv:2412.16720}.

\bibitem[{Jain et~al.(2024)Jain, Han, Gu, Li, Yan, Zhang, Wang, Solar-Lezama, Sen, and Stoica}]{livecodebench}
Naman Jain, King Han, Alex Gu, Wen-Ding Li, Fanjia Yan, Tianjun Zhang, Sida Wang, Armando Solar-Lezama, Koushik Sen, and Ion Stoica. 2024.
\newblock Livecodebench: Holistic and contamination free evaluation of large language models for code.
\newblock \emph{arXiv preprint arXiv:2403.07974}.

\bibitem[{Jimenez et~al.(2023)Jimenez, Yang, Wettig, Yao, Pei, Press, and Narasimhan}]{swe-bench}
Carlos~E Jimenez, John Yang, Alexander Wettig, Shunyu Yao, Kexin Pei, Ofir Press, and Karthik Narasimhan. 2023.
\newblock Swe-bench: Can language models resolve real-world github issues?
\newblock \emph{arXiv preprint arXiv:2310.06770}.

\bibitem[{Lai et~al.(2023)Lai, Li, Wang, Zhang, Zhong, Zettlemoyer, Yih, Fried, Wang, and Yu}]{ds-1000}
Yuhang Lai, Chengxi Li, Yiming Wang, Tianyi Zhang, Ruiqi Zhong, Luke Zettlemoyer, Wen-tau Yih, Daniel Fried, Sida Wang, and Tao Yu. 2023.
\newblock Ds-1000: A natural and reliable benchmark for data science code generation.
\newblock In \emph{International Conference on Machine Learning}, pages 18319--18345. PMLR.

\bibitem[{Li et~al.(2023)Li, Allal, Zi, Muennighoff, Kocetkov, Mou, Marone, Akiki, Li, Chim et~al.}]{starcoder}
Raymond Li, Loubna~Ben Allal, Yangtian Zi, Niklas Muennighoff, Denis Kocetkov, Chenghao Mou, Marc Marone, Christopher Akiki, Jia Li, Jenny Chim, and 1 others. 2023.
\newblock Starcoder: may the source be with you!
\newblock \emph{arXiv preprint arXiv:2305.06161}.

\bibitem[{Li et~al.(2025)Li, Dong, Lee, Xia, Zhang, Dai, Wang, and Tang}]{li-etal-2025-coir}
Xiangyang Li, Kuicai Dong, Yi~Quan Lee, Wei Xia, Hao Zhang, Xinyi Dai, Yasheng Wang, and Ruiming Tang. 2025.
\newblock \href {https://doi.org/10.18653/v1/2025.acl-long.1072} {{C}o{IR}: A comprehensive benchmark for code information retrieval models}.
\newblock In \emph{Proceedings of the 63rd Annual Meeting of the Association for Computational Linguistics (Volume 1: Long Papers)}, pages 22074--22091, Vienna, Austria. Association for Computational Linguistics.

\bibitem[{Li et~al.(2022)Li, Choi, Chung, Kushman, Schrittwieser, Leblond, Eccles, Keeling, Gimeno, Dal~Lago et~al.}]{codecontest}
Yujia Li, David Choi, Junyoung Chung, Nate Kushman, Julian Schrittwieser, R{\'e}mi Leblond, Tom Eccles, James Keeling, Felix Gimeno, Agustin Dal~Lago, and 1 others. 2022.
\newblock Competition-level code generation with alphacode.
\newblock \emph{Science}, 378(6624):1092--1097.

\bibitem[{Liu et~al.(2024)Liu, Feng, Xue, Wang, Wu, Lu, Zhao, Deng, Zhang, Ruan et~al.}]{deepseek}
Aixin Liu, Bei Feng, Bing Xue, Bingxuan Wang, Bochao Wu, Chengda Lu, Chenggang Zhao, Chengqi Deng, Chenyu Zhang, Chong Ruan, and 1 others. 2024.
\newblock Deepseek-v3 technical report.
\newblock \emph{arXiv preprint arXiv:2412.19437}.

\bibitem[{Microsoft(2024)}]{markitdown}
Microsoft. 2024.
\newblock \href {https://github.com/microsoft/markitdown} {Python tool for converting files and office documents to markdown.}

\bibitem[{Muennighoff et~al.(2025)Muennighoff, Yang, Shi, Li, Fei-Fei, Hajishirzi, Zettlemoyer, Liang, Cand{\`e}s, and Hashimoto}]{s1}
Niklas Muennighoff, Zitong Yang, Weijia Shi, Xiang~Lisa Li, Li~Fei-Fei, Hannaneh Hajishirzi, Luke Zettlemoyer, Percy Liang, Emmanuel Cand{\`e}s, and Tatsunori Hashimoto. 2025.
\newblock s1: Simple test-time scaling.
\newblock \emph{arXiv preprint arXiv:2501.19393}.

\bibitem[{{OpenAI}(2025)}]{o4-mini}
{OpenAI}. 2025.
\newblock \href {https://openai.com/index/introducing-o3-and-o4-mini/} {Introducing o3 and o4-mini}.

\bibitem[{Roziere et~al.(2023)Roziere, Gehring, Gloeckle, Sootla, Gat, Tan, Adi, Liu, Sauvestre, Remez et~al.}]{codellama}
Baptiste Roziere, Jonas Gehring, Fabian Gloeckle, Sten Sootla, Itai Gat, Xiaoqing~Ellen Tan, Yossi Adi, Jingyu Liu, Romain Sauvestre, Tal Remez, and 1 others. 2023.
\newblock Code llama: Open foundation models for code.
\newblock \emph{arXiv preprint arXiv:2308.12950}.

\bibitem[{Yang et~al.(2025{\natexlab{a}})Yang, Li, Yang, Zhang, Hui, Zheng, Yu, Gao, Huang, Lv et~al.}]{qwen3}
An~Yang, Anfeng Li, Baosong Yang, Beichen Zhang, Binyuan Hui, Bo~Zheng, Bowen Yu, Chang Gao, Chengen Huang, Chenxu Lv, and 1 others. 2025{\natexlab{a}}.
\newblock Qwen3 technical report.
\newblock \emph{arXiv preprint arXiv:2505.09388}.

\bibitem[{Yang et~al.(2025{\natexlab{b}})Yang, Huang, Zhang, Surdeanu, Wang, and Pan}]{yang2025llmreason}
Minglai Yang, Ethan Huang, Liang Zhang, Mihai Surdeanu, William Wang, and Liangming Pan. 2025{\natexlab{b}}.
\newblock \href {https://arxiv.org/abs/2505.18761} {How is llm reasoning distracted by irrelevant context? an analysis using a controlled benchmark}.
\newblock \emph{Preprint}, arXiv:2505.18761.

\bibitem[{Zhu et~al.(2024)Zhu, Guo, Shao, Yang, Wang, Xu, Wu, Li, Gao, Ma et~al.}]{deepseek-coder-v2}
Qihao Zhu, Daya Guo, Zhihong Shao, Dejian Yang, Peiyi Wang, Runxin Xu, Y~Wu, Yukun Li, Huazuo Gao, Shirong Ma, and 1 others. 2024.
\newblock Deepseek-coder-v2: Breaking the barrier of closed-source models in code intelligence.
\newblock \emph{arXiv preprint arXiv:2406.11931}.

\end{thebibliography}
